\documentclass[conference]{IEEEtran}
\usepackage{subfigure}
\IEEEoverridecommandlockouts
\usepackage{cite}
\usepackage{amsmath,amssymb,amsfonts}
\usepackage{algorithmic}
\usepackage{graphicx}
\usepackage{textcomp}
\usepackage{xcolor}
\def\BibTeX{{\rm B\kern-.05em{\sc i\kern-.025em b}\kern-.08em
    T\kern-.1667em\lower.7ex\hbox{E}\kern-.125emX}}
\newcommand\blfootnote[1]{%
  \begingroup
  \renewcommand\thefootnote{}\footnote{#1}%
  \addtocounter{footnote}{-1}%
  \endgroup
}

\DeclareRobustCommand*{\IEEEauthorrefmark}[1]{%
\raisebox{0pt}[0pt][0pt]{\textsuperscript{\footnotesize\ensuremath{#1}}}}

\begin{document}

\title{Transformer-based Learned Image Compression for Joint Decoding and Denoising\\
}

\author{%
\IEEEauthorblockN{
Yi-Hsin Chen\IEEEauthorrefmark{1} 
\quad Kuan-Wei Ho\IEEEauthorrefmark{1} 
\quad Shiau-Rung Tsai\IEEEauthorrefmark{1} 
\quad Guan-Hsun Lin\IEEEauthorrefmark{1} \\
\quad Alessandro Gnutti\IEEEauthorrefmark{2}
\quad Wen-Hsiao Peng\IEEEauthorrefmark{1}
\quad Riccardo Leonardi\IEEEauthorrefmark{2}%
}
\IEEEauthorblockA{%
\IEEEauthorrefmark{1}National Yang Ming Chiao Tung University, Taiwan \quad \IEEEauthorrefmark{2}University of Brescia, Italy}
%\IEEEauthorblockA{%
%\tt\small \{yhchen12101.cs09@, kwho@cs., mick20001108.cs12@, abc900203abc.cs12@\}nycu.edu.tw \\ \tt\small wpeng@cs.nctu.edu.tw \tt\small alessandro.gnutti@unibs.it}
}

% {yhchen12101.cs09, jefferyho.cs12, mick20001108.cs12, abc900203abc.cs12\}@nycu.edu.tw

% \and
% \and
% \and
% \and
% \and

\maketitle

\begin{abstract}
This work introduces a Transformer-based image compression system. It has the flexibility to switch between the standard image reconstruction and the denoising reconstruction from a single compressed bitstream. Instead of training separate decoders for these tasks, we incorporate two add-on modules to adapt a pre-trained image decoder from performing the standard image reconstruction to joint decoding and denoising. Our scheme adopts a two-pronged approach. It features a latent refinement module to refine the latent representation of a noisy input image for reconstructing a noise-free image. Additionally, it incorporates an instance-specific prompt generator that adapts the decoding process to improve on the latent refinement. Experimental results show that our method achieves a similar level of denoising quality to training a separate decoder for joint decoding and denoising at the expense of only a modest increase in the decoder's model size and computational complexity.

\end{abstract}

\vspace{-0.2cm}
\blfootnote{\textcolor{black}{This work was supported by National Science and Technology Council, Taiwan, under Grants NSTC-112-2634-F-A49-007- and MOST-110-2221-E-A49-065-MY3, National Center for High-performance Computing, Taiwan, and partially supported by the European Union under the Italian National Recovery and Resilience Plan (NRRP) of NextGenerationEU, partnership on “Telecommunications of the Future” (PE00000001 - program “RESTART”).}}

\section{Introduction}
\label{sec:intro}
Natural images captured by digital imaging sensors often exhibit noise due to sensor limitations, ISO settings, low-light conditions, etc. To transmit a noisy image efficiently, a straightforward approach is to perform pre-processing (or post-processing) to remove the noise before (or after) compression. Recent studies~\cite{optimizingIC2022, brummer_wacv23} show that, as compared to straightforwardly cascading an image codec and a denoising model, optimizing a learned image codec for joint decoding and denoising is more effective in terms of both compression performance and complexity. \cite{optimizingIC2022, brummer_wacv23} use clean-noise image pairs to optimize the entire image codec end-to-end with a rate-distortion loss. With this approach, the noise is filtered out to a large extent in the encoder. This results in the original noisy input image not being able to be recovered on the decoder side.  

However, there are applications where the preservation of image noise is crucial to the trustworthiness of images. Examples include medical imaging, satellite imagery, and security surveillance, just to name a few. In 2022, the JPEG standard committee launched a JPEG AI project, issuing a call for proposals~\cite{jpegaicfp} with the aim of standardizing a learning-based image codec. This codec is designed to transmit a single bitstream that is catered to both the standard image reconstruction and the compressed-domain image processing and machine tasks. In this context, Larigauderie~\emph{et~al.}~\cite{larigauderie2022combining} propose to optimize the decoder exclusively for joint decoding and denoising, given that the encoder is pre-trained for the standard image reconstruction. Taking a different approach, Alvar~\emph{et~al.}~\cite{ranjbar2022joint} introduce a scalable coding scheme, allowing the clean image to be decoded from a subset of the compressed latents. To reconstruct the noisy input image, the entire latents needs to be decoded. Both \cite{larigauderie2022combining} and \cite{ranjbar2022joint} require separate decoders to switch between the standard image reconstruction and denoising reconstruction.

\begin{figure}[t]
    \centering
    \includegraphics[width=.99\linewidth]{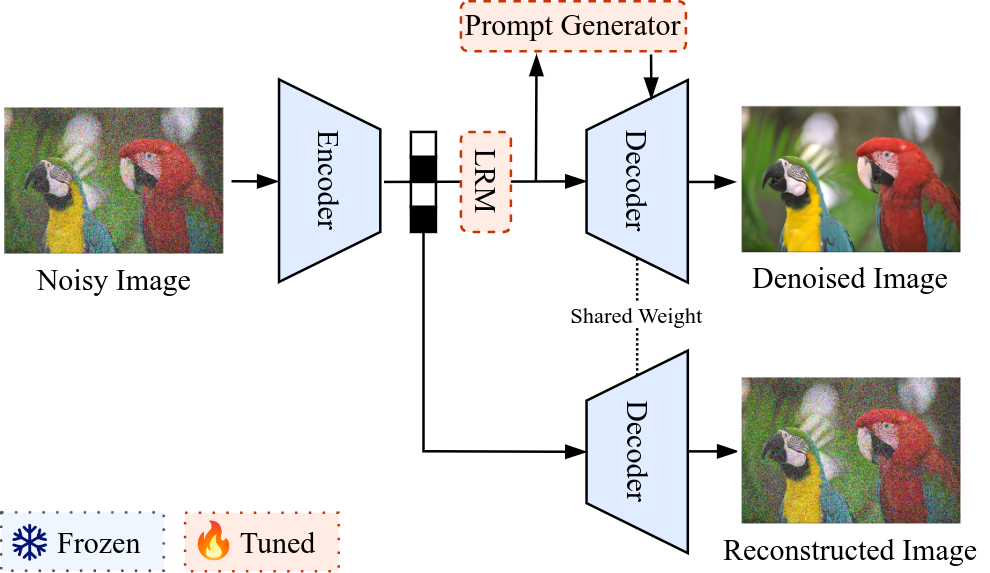}
    \vspace{-0.6cm}
    \caption{Overview of the proposed joint decoding and denoising framework.}
    \label{fig:teaser}
    \vspace{-0.5cm}
\end{figure}
%Among them, the standard reconstruction is a mandatory task, while image processing and machine tasks are considered secondary. 

\begin{figure*}[t!]
    \centering
    \includegraphics[width=\linewidth]{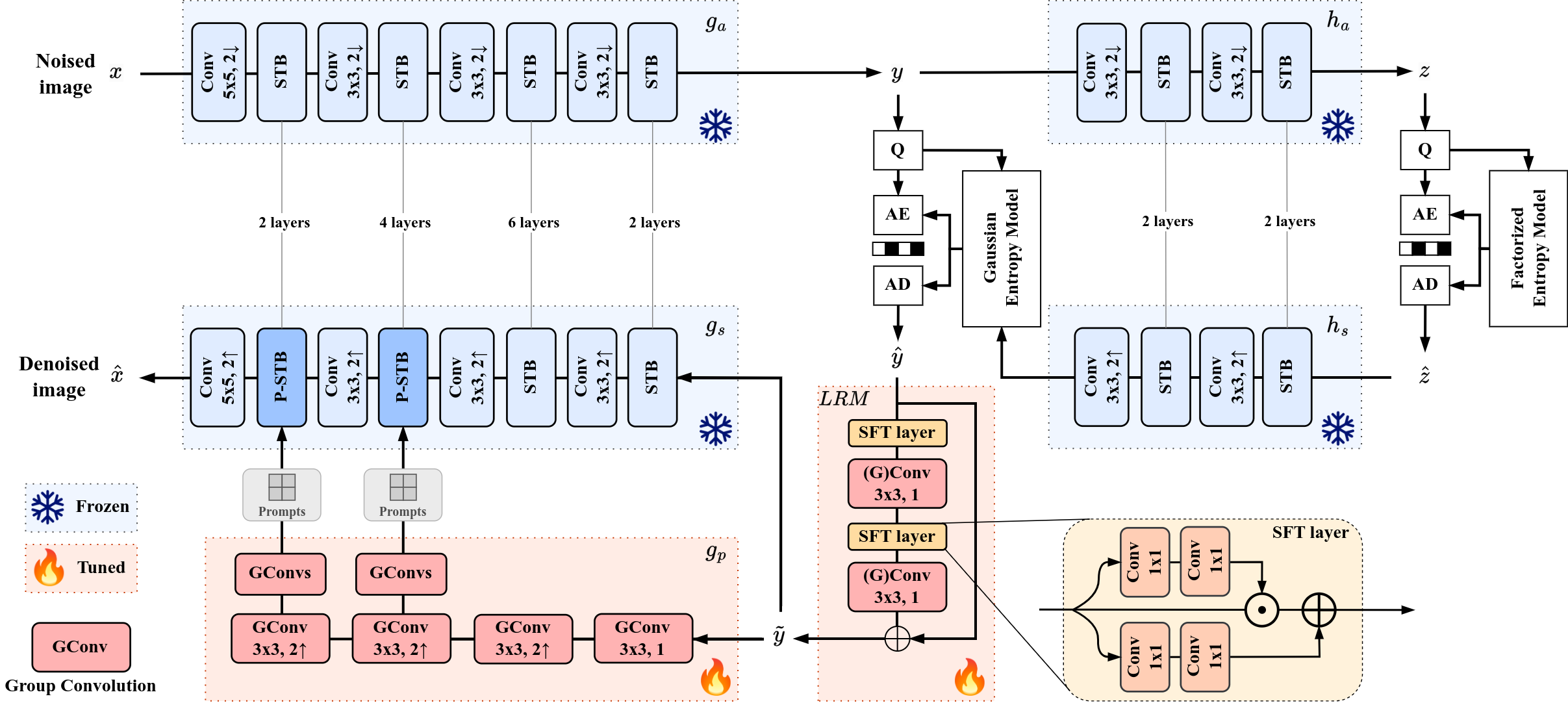}
    \vspace{-0.6cm}
    \caption{Illustration of the proposed method. Our work adapts a pre-trained Transformer-based image codec to perform joint image decoding and denoising by introducing a latent refinement module LRM and a prompt generator $g_p$ on the decoder side. The GConv (highlighted in pink color) represents sixteen groups of $3 \times 3$ group convolutions. \textcolor{black}{The (G)Conv in LRM is convolutions for our proposed model, and group convolutions for its lightweight variant.} Although the picture does not explicitly show it for better visualization, our system can directly process the latent $\hat{y}$ for standard image reconstruction.}
    \label{fig:main}
    \vspace{-0.4cm}
\end{figure*}

Our work proposes a Transformer-based image compression system that allows the user to switch between the standard image reconstruction and denoising reconstruction from a single compressed bitstream at inference time. As depicted in Fig.~\ref{fig:teaser}, our approach starts with a pre-trained base codec optimized for the standard image reconstruction. When the latents of a noisy input image are decoded, the decoder produces a noisy reconstruction of the input image. To perform joint decoding and denoising, we introduce two add-on modules, Latent Refinement Module (LRM) and Prompt Generator. The former aims to predict the latent representation of a clean image from that of a noisy input image, while the latter is designed to adapt the decoding process of the base decoder to suit the needs of joint decoding and denoising. Our add-on approach allows the base decoder to be re-used and re-purposed without retraining. It incurs only a marginal 28\% increase in the decoder's model size, as compared to training a separate decoder (i.e. a 100\% increase in the model size) dedicated to the denoising reconstruction. Furthermore, in terms of the number of the multiply–accumulate operations, our prompt-adapted decoder is only 11\% higher than the base decoder.

\section{Proposed Method}
\label{sec:method}
\subsection{System Overview}
\label{Sec:Overview}

This work aims at adapting a pre-trained Transformer-based image codec optimized for the image reconstruction task to perform joint image denoising and decoding. It provides the user with an option to choose between the standard image reconstruction and the denoising reconstruction from a single compressed bitstream. To this end, we introduce a decoder-side adaptation mechanism capable of re-purposing the pre-trained decoder according to the user's preference at inference time.

Fig.~\ref{fig:main} illustrates our proposed scheme. We adopt the same pre-trained base codec (highlighted in blue and white colors) as~\cite{tic} \textcolor{black}{but replace the complicated context model with simple Gaussian prior for entropy coding}. In Fig.~\ref{fig:main}, $g_a, g_s$ and $h_a, h_s$ represent the main and hyperprior autoencoders, respectively. The main autoencoder $g_a, g_s$ consists of multiple Swin-Transformer blocks (STBs), with the convolutional and de-convolutional layers interspersed between STBs to adjust the resolution of the feature maps. Since our base image codec is pre-trained for the standard image reconstruction task, a noisy input image leads to a noisy reconstructed image at the decoder's output. To reuse this pre-trained base codec for compressed-domain image denoising, we introduce several add-on modules, including an LRM and a prompt generator $g_p$, on the decoder side while leaving the base codec untouched. Notably, given a compressed bitstream, our scheme supports both the standard image reconstruction task and the joint decoding and denoising task. \textcolor{black}{The add-on modules represent a fractional increase in the decoder's model size and computational complexity.}

%it reconstructs a noisy output image when the coded latents $\hat{y}$ of the noisy input image is decoded by the decoder $g_s$.  

\subsection{Latent Refinement Module}
\label{Sec:LRM}
The LRM is to refine the latent representation $\hat{y}$ of a noisy input image, with the aim of predicting the latents of the clean (denoised) image. When the prediction is perfect, the pre-trained decoder should ideally produce a clean reconstruction of the input image, i.e. a denoised image. In a sense, LRM performs compressed-domain image denoising. 

As depicted in Fig~\ref{fig:main}, LRM is a residual block in which the residual generation involves two Spatial Feature Transforms (SFT)~\cite{sft} interleaved with two $3 \times 3$ convolutional layers. The SFT applies spatially adaptive affine transformation, in order to update the feature maps $F$ by $\text{SFT}(F)= (\alpha(F) \odot F) \oplus \beta (F)$, where $\odot$ and $\oplus$ denote element-wise multiplication and addition, respectively. Both $\alpha(\cdot)$ and $\beta(\cdot)$ are learned neural networks. In an effort to reduce the model size and computational cost of LRM, we also explore an alternative implementation of LRM that replaces $3 \times 3$ convolutions with sixteen groups of $3 \times 3$ group convolutions\textcolor{black}{\cite{groupconv-alexnet}}. This is referred hereafter to as our lightweight variant. 

%introduced by this add-on module LRM during the entire decoding process

%Note that the pre-trained decoder is trained to decode a coded latent back to an image without performing denoising.
%we first refine the noisy latent $\hat{y}$ before feeding it to the pre-trained decoder. 

\begin{figure}[t]
    \centering
    \includegraphics[width=0.98\linewidth]{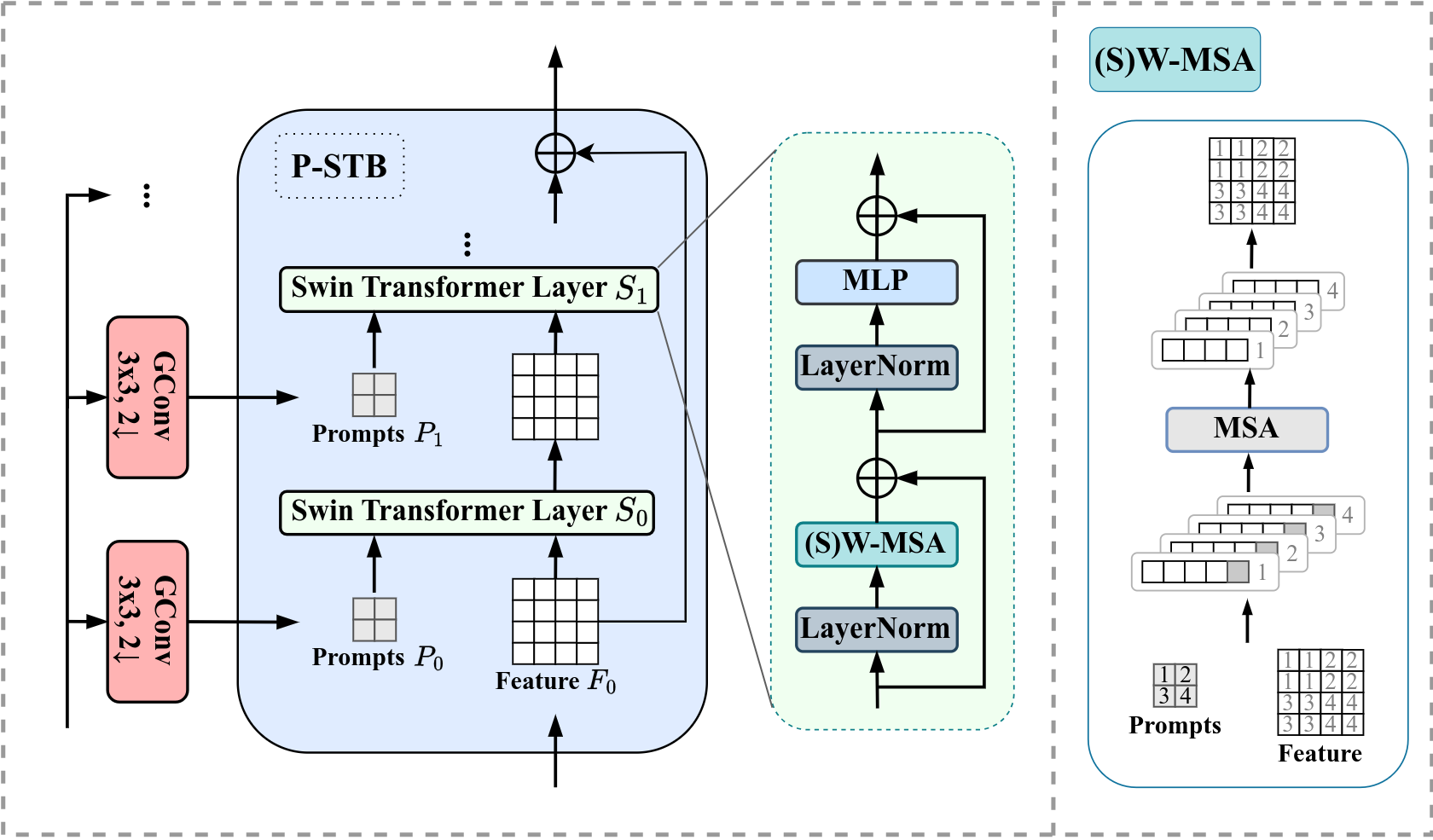}
    \vspace{-0.2cm}
    \caption{Network details of the prompt-adapted Swin-Transformer Layer. }
    \label{fig:STB}
    \vspace{-0.4cm}
\end{figure}

\subsection{Prompt-adapted Swin-Transformer}
Recognizing that the predictions made by LRM are not perfect and our Transformer-based base codec is pre-trained and frozen for the standard image reconstruction, we propose a decoder-side, instance-specific prompting technique to adapt the decoder to suit the needs of joint decoding and denoising. Specifically, additional tokens, known as prompts, are generated and injected into the STBs of the decoder without re-training the decoder.

As shown in Fig~\ref{fig:main}, our prompt generator $g_p$ consists of several $3 \times 3$ group convolutions. It takes the refined latents $\tilde{y}$ to generate prompts for the last two STBs (closer to the input image space) in the decoder. These prompts change with $\tilde{y}$ and are thus instance-specific (i.e. image-dependent). Fig.~\ref{fig:STB} illustrates the structure of our prompt-adapted Swin-Transformer block (P-STB). The P-STB comprises several Swin-Transformer layers~\cite{swintransformer}, with shifted window-based multi-head self-attention (W-MSA) serving as the main mechanism for signal transformation. In the absence of prompts, P-STB is an ordinary STB, where W-MSA divides the input tokens $F$ into non-overlapping windows (groups). The tokens in every window are updated through self-attention. In symbols, the attention mechanism is given by
\begin{align}
    \text{Attention}(Q,K,V) &= \text{Softmax}(QK^\top/\sqrt{d}+B)V,
    \label{eq:attention}
\end{align}
where $Q = FW_Q$, $K = FW_K$, $V = FW_V$, $W_Q, W_K, W_V\in\mathbb{R}^{C\times C}$ are learnable matrices mapping flattened input $F\in\mathbb{R}^{N \times C}$ into query $Q\in\mathbb{R}^{N\times C}$, key $K\in\mathbb{R}^{N\times C}$, and value $V\in\mathbb{R}^{N\times C}$. Here, $N$ is the number of input tokens in a window, $C$ is the channel dimension of $F$, and $B \in \mathbb{R}^{N\times N}$ is a learnable positional embedding matrix. 

When the prompts $P$ are present, they are spatially divided in the same way as input tokens $F$. In our design, the number of prompts is one-quarter of that of the input tokens to reduce complexity. In a self-attention window, prompts from the corresponding window update the input tokens following a similar self-attention mechanism to Eq.~\eqref{eq:attention}, except that the key $K$ and value $V$ matrices are augmented as $K = [F; P]W_K \in\mathbb{R}^{(N+\frac{N}{4})\times C}$, $V = [F;P] W_V\in\mathbb{R}^{(N+\frac{N}{4})\times C}$, where $[\cdot;\cdot]$ represents concatenation along the token dimension, while query $Q = FW_Q$ remain unchanged. This prompting technique allows the decoding process to be adapted without retraining the base decoder.

\begin{figure*}[t]
\centering
\subfigure[Real-world noise]{
    \centering
    \includegraphics[width=.32\textwidth]{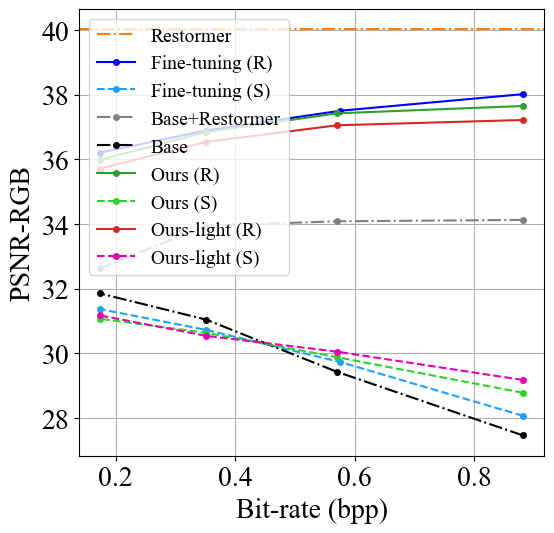}
    \label{fig:rd-a}
    }
\hspace{-0.35cm}
\subfigure[Synthetic noise]{
    \centering
    \includegraphics[width=.32\textwidth]{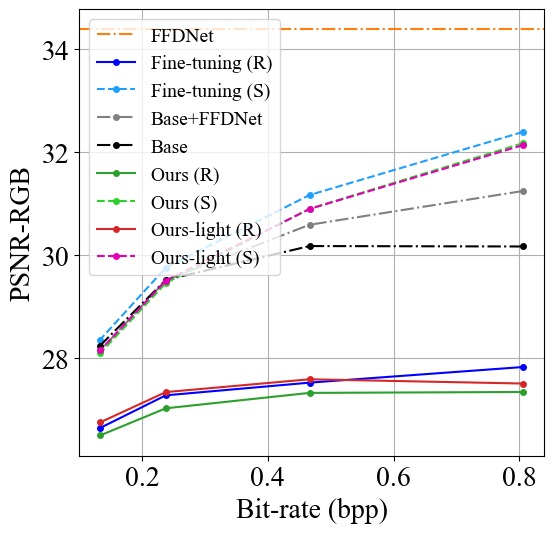}
    \label{fig:rd-b}
}
\hspace{-0.35cm}
\subfigure[Ablation]{
    \centering
    \includegraphics[width=.32\textwidth]{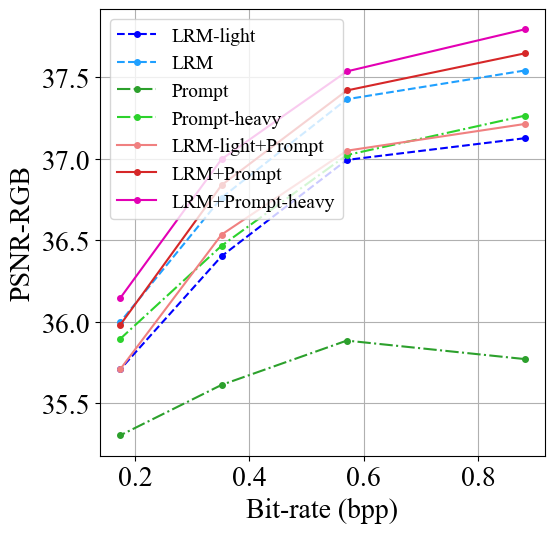}
    \label{fig:rd-c}
}
\vspace{-0.2cm}
\caption{The rate-distortion comparison of the competing methods tested for (a) real-world noise and (b) synthetic noise. Restormer and FFDNet are trained on real-world noise and synthetic noise, respectively. \textcolor{black}{(R) and (S) indicate the models trained on real-world noise and synthetic noise, respectively.} (c) shows the ablation experiment \textcolor{black}{on real-world noise.}}% Prompt is the proposed method and Prompt-L is its heavy variant.}}
\label{fig:RD}
\vspace{-0.4cm}
\end{figure*}

% (a), (b) are the RD-curve comparison with our model and baseline methods. Restormer and FFDNet are trained on real-world noise and synthetic noise, respectively. Fine-tuning (R) means the Base model trained on the decoder side for Real-world noise, and Fine-tuning (S) means for Synthetic noise. (c) shows the performance of ablation experiment. Prompt-L has the same model size as LRM, and Prompt is what we used in our work.

\section{Experiments}

\label{sec:experiment}

%\subsection{Experimental Settings}
{\bf Training details:} 
Following the common test protocol~\cite{optimizingIC2022}, we test our model for both real-world and synthetic noise scenarios. For the real-world case, we train and validate our model on SIDD dataset~\cite{sidd}, which has 320 high-resolution images and 1280 256x256 patches. For the synthetic case, we use Flicker2W dataset~\cite{flicker2w} for training and Urban100~\cite{urban100} dataset for validation. To construct clean-noise image pairs, we use the same noise simulator~\cite{simulator} as adopted by JPEG AI~\cite{jpeg-ai}. It is optimized for estimating the parameters of Poissonian-Gaussian noise model on SIDD dataset~\cite{sidd}. 

We adopt a two-stage training strategy. In the first stage, the base codec $g_a, g_s, h_a, h_s$ is trained on Flicker2W~\cite{flicker2w} for the standard image reconstruction task. The training objective is to minimize the rate-distortion cost $R(\hat{z})+R(\hat{y}) + \lambda \times D ( x, \hat{x})$, where $R$ denotes the bit rates of the image latents and hyperprior, and $D$ measures the mean-squared error between the input $x$ and reconstructed $\hat{x}$ images. $\lambda$ is set to 0.0018, 0.0035, 0.0067, and 0.013 for separate rate points. In the second stage, we fix the base codec $g_a, g_s, h_a, h_s$, and train \textcolor{black}{4 pairs of} LRM and the prompt generator $g_p$ \textcolor{black}{for 4 distinct rate points} by \textcolor{black}{following the common training strategy for the image restoration task to minimize} the $l_1$ loss between \textcolor{black}{the clean image $x$ and the denoising reconstruction $\hat{x}$}.

%Follow [paper], we perform denoising experiments on both real-world and synthetic noise datasets. For real-world noise, we use SIDD training and valadition dataset, which compose of 320 high resolution image and 1280 patches of 256x256 respectively. For the experiments in synthetic noise, we use Flicker2W dataset for training and Urban100 dataset for validation. There are 20,745 general clean images in the Flicker2W dataset and 100 images of urban scenes in the Urban100 dataset. About synthetic noise, we use the practical noise simulator from [paper]. The practical noise simulator was created by fitting Poissonian-Gaussian noise model [paper] to the noise from SIDD dataset [ref] and it is also utilized in the JPEG AI standardation for the evaluation of image denoising task. During training step, we use image patches randomly cropped at a resolution of 256 pixels to optimized our model. Our training objective can be formulated as Loss = $\lambda \times L1 ( \hat{x}, x) $.

{\bf Evaluation:}
We test our model on SIDD test set~\cite{sidd} for real-world noise, and Kodak~\cite{kodak} for synthetic noise, which is generated with the same noise simulator~\cite{simulator} as that for training. The quality of denoised images is measured in PSNR-RGB with the noise-free image serving as the original, and the bit-rate in bits-per-pixel (bpp).

%We evaluate our model performance on SIDD test dataset for real-world noise, and on Kodak24 dataset for synthetic noise. We adapt the peak signal-to-noise ratio (PSNR) as the quality metric for both real-world and synthetic noise experiments. 

{\bf Baselines:}
%As our work allows the user to switch between the standard image reconstruction and the denoising reconstruction on the decoder side, our comparison is focused on the baseline methods that perform denoising on the decoder side only. 
The baseline methods include (1) using the base codec $(g_a, g_s, h_a, h_s)$ trained for the standard image reconstruction, termed \textit{Base}, (2) fine-tuning the main decoder $g_s$ of the base codec, termed \textit{Fine-tuning}, and (3) using a pre-trained denoising model for post-processing, denoted as \textit{Base + model\_name}, e.g.~\textit{Base + Restormer}. We adopt Restormer~\cite{restormer} for real-world noise and FFDNet~\cite{ffdnet} for synthetic noise. Both are the state-of-the-art models. Recall that our work allows the user to switch between the standard image reconstruction and the denoising reconstruction. The decoder is still allowed to reconstruct the noisy input image from the compressed bitstream. As such, we exclude the pre-processing approach, namely denoising followed by compression, for a fair comparison. 
%We employed three baseline methods,including the base codec (TIC), fine-tuning the base codec on the decoder side, and combining a denoising model following TIC. The denoise models are Restormer[??] trained on SIDD dataset for real-world noise and FFDNet[??] trained on Poisson-Gaussian distribution for synthetic noise. Restormer is robust denoising model, and FFDNet is an anchor in JPEG AI[??].

\subsection{Rate-Distortion Performance}
%\subsection{Rate-Distortion and Subjective Quality Comparison}
Figs.~\ref{fig:RD}(a) and \ref{fig:RD}(b) compare the rate-distortion performance of the competing methods for real-world and synthesis noise, respectively. \textcolor{black}{The terms (R) and (S) indicate the models trained on real-world and synthetic noise, respectively. From the figure,} we make the following observations. (1) First, both the normal and lightweight versions of our proposed method achieve substantial gains over \textit{Base}, which simply performs image reconstruction without denoising. This showcases the effectiveness of our approach in performing joint decoding and denoising without retraining the base codec. (2) Second, both our method and \textit{Fine-tuning} outperform the post-processing approaches, \textit{Base+Restormer} and \textit{Base+FFDNet}. Note that these approaches cascade two pre-trained models, forming collectively a large compound model. (3) Third, our method shows comparable or slightly inferior performance to \textit{Fine-tuning}. However, when trained for a specific noise distribution and tested under a different noise distribution (e.g. \textit{Fine-tuning~(S)} in Fig.~\ref{fig:RD}(a) and \textit{Fine-tuning~(R)} in Fig.~\ref{fig:RD}(b)), the performance of \textit{Fine-tuning} deteriorates significantly. This indicates the poor generalization of \textit{Fine-tuning} to different noise distributions. This limitation calls for the need to train separate decoders for drastically different noise distributions. Although suffering from similar generalization issues, our approach only needs to update the lightweight add-on modules, i.e. LRM and the prompt generator $g_p$, while reusing the base decoder completely. It offers a more cost-effective solution than \textit{Fine-tuning}.

\begin{figure*}[t]
\centering
\subfigure[Real-world noise]{
    \centering
    \includegraphics[width=.97\linewidth]{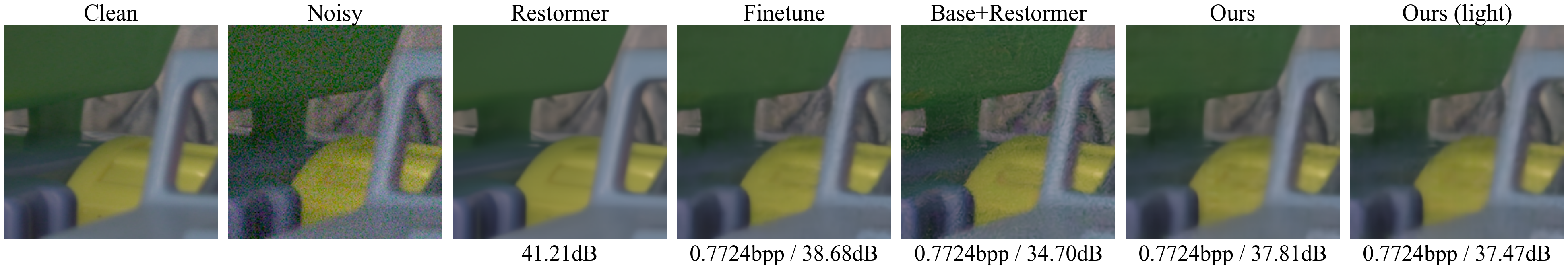}
    \label{fig:vis-a}
    }
\subfigure[Synthetic noise]{
    \centering
    \includegraphics[width=.97\linewidth]{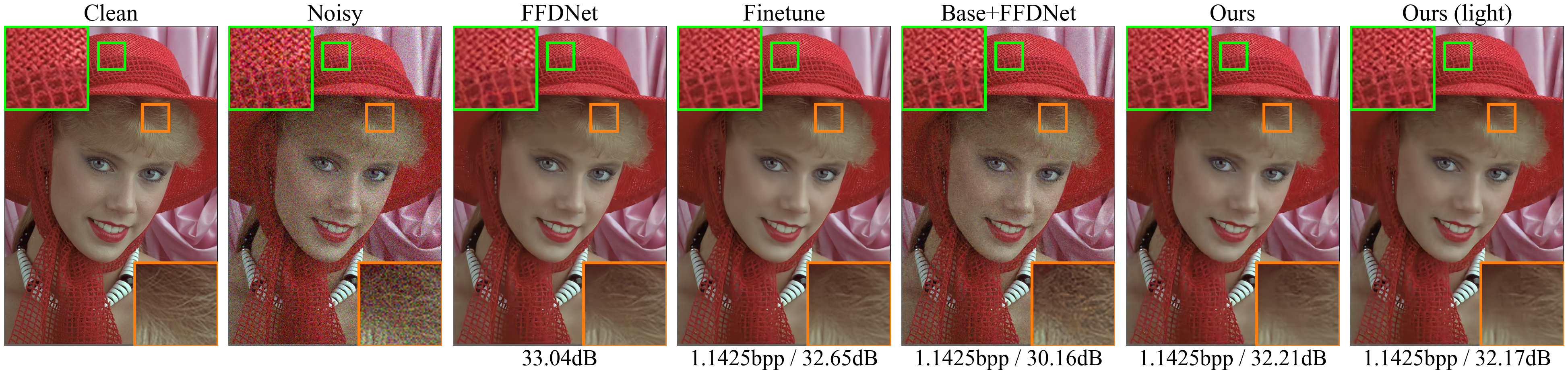}
    \label{fig:vis-b}
}
\vspace{-0.2cm}
\caption{Visualization of the noise-free, noisy, and denoised images. The top row are results with real-world noise. The second row are results with synthetic noise. Bits-per-pixel (bpp) / PSNR-RGB are provided for comparison. Zoom in for better visualization.}
\label{fig:visualization}
\vspace{-0.4cm}
\end{figure*}

Fig.~\ref{fig:visualization} further presents the subjective quality comparison. Both our method and \textit{Fine-tuning} are able to denoise the input image to a large extent. In comparison, the post-processing methods exhibit rather noticeable noise or artifacts. It is expected that training these compound models end-to-end should improve the performance at the cost of high complexity.

\subsection{Complexity Comparison}
\label{sec:complexity}
% \begin{table}[]
% \centering
% \caption{}
% \setlength{\tabcolsep}{5.0pt}
% \renewcommand{\arraystretch}{1.2}
% \begin{tabular}{cl|c|c}
% \hline
% \multicolumn{1}{l}{}      &                 & {\textbf{kMACs/pixel}}    & {\textbf{Params(M)}} \\ \hline
% \multicolumn{1}{l}{}      & TIC             & 188.52          & 3.87                \\ \hline
% \multicolumn{1}{l}{}      & Ours            & 209.53 (+11\%)  & 4.94 (+28\%)         \\
% \multicolumn{1}{l}{}      & Ours (light)    & 207.10 (+10\%)  & 4.32 (+12\%)         \\ \hline
% \multicolumn{1}{c}{ablation} & SFT          & 192.27 (+ 2\%)  & 4.83 (+25\%)         \\
% \multicolumn{1}{c}{}         & Prompt       & 205.78 (+ 9\%)  & 3.98 (+ 3\%)         \\
% \multicolumn{1}{c}{}         & Prompt-L     & 255.16 (+35\%)  & 4.83 (+25\%)         \\ \hline
% \multicolumn{1}{c}{Prompts}  & Prompt-12    & 197.55 (+ 5\%)  & 4.98 (+29\%)         \\
% \multicolumn{1}{c}{}         & Prompt-1234  & 211.59 (+12\%)  & 5.04 (+30\%)         \\ \hline
% \end{tabular}
% \end{table}

\begin{table}[]
\centering
\caption{Comparison of the kMACs/pixel and model size.}
\label{tab:main_complexity}
\setlength{\tabcolsep}{5.0pt}
\renewcommand{\arraystretch}{1.2}
\begin{tabular}{l|c|c}
\hline
                & {\textbf{kMACs/pixel}}  & {\textbf{Params (M)}} \\ \hline
Base            & 188.52                  & 3.87                 \\ \hline
Fine-tuning     & 188.52 (+ 0\%)          & 7.74 (+100\%)         \\ 
Base+Restormer  & 1461.10 (+675\%)  & 28.40 (+634\%)   \\ 
Base+FFDNet     & 402.72 (+114\%)   & 4.72 (+22\%)    \\ 
Ours            & 209.53 (+11\%)          & 4.94 (+28\%)         \\
Ours (light)    & 207.10 (+10\%)          & 4.32 (+12\%)         \\ \hline
\end{tabular}
\vspace{-0.45cm}
\end{table}
Table~\ref{tab:main_complexity} reports the decoder-side complexity of the competing methods in terms of (1) the required model size for providing the user with the option to switch between the standard image reconstruction and denoising reconstruction and (2) the number of kilo-multiply-accumulate-operations (kMACs/pixel) for the denoising reconstruction. 

Compared with \textit{Fine-tuning}, which requires one separate full decoder (100\%) for the denoising reconstruction, our method incurs only a 28\% increase in model size. Remarkably, our lightweight variant has a marginal 12\% increase in model size while showing a modest impact on the image quality. Both of these variants increase the kMACs/pixel by about 10\% relative to a full decoder for the denoising reconstruction. Note that the post-processing approaches are least preferred due to their high kMACs/pixel.

%In contrast, while our method and \textit{fine-tuning} directly apply denoising in the latent domain, post-processing methods need to reconstruct back to the image domain and then use another network to re-extract features for denoising, resulting in more MAC operations.

%Table 1. compares our model with the \textit{Base} in terms of complexity measured by kilo-multiply-accumulate-operations per pixel (kMACs/pixel) and model size. Our model requires an increase of only 28\% in parameters and 11\% in kMACs/pixel compared to the \textit{Base} on the decoder side. However, ours method using a lightweight SFT module, which has little effect on performance, results only a 12\% increase in parameters and an 11\% increase in kMACs/pixel, respectively. Note that our model effectively performs denoising tasks in the compressed domain with training only a few number of additional parameters.

%\input{Figure/ablation_RD-curve}

\subsection{Ablation Experiment}
\begin{table}[]
\centering
\caption{Complexity comparison between \textcolor{black}{different variants of LRMs and the prompt generators.}}
\label{tab:ablation_complexity}
\setlength{\tabcolsep}{5.0pt}
\renewcommand{\arraystretch}{1.2}
\begin{tabular}{l|c|c}
\hline
           & {\textbf{kMACs/pixel}}  & {\textbf{Params (M)}} \\ \hline
Base       & 188.52                  & 3.87                 \\ \hline
LRM-light & 189.84 (+ 1\%)          & 4.21 (+ 9\%)         \\ 
LRM        & 192.27 (+ 2\%)          & 4.83 (+25\%)         \\ 
Prompt     & 205.78 (+ 9\%)          & 3.98 (+ 3\%)         \\ 
Prompt-heavy   & 285.26 (+51\%)          & 7.15 (+84\%)         \\ 
% Prompt-L   & \textcolor{red}{255.16 (+35\%)}          & \textcolor{red}{4.83 (+25\%)}         \\ 
\textbf{LRM-light+Prompt} & 207.10 (+10\%)  & 4.32 (+12\%) \\
\textbf{LRM+Prompt} & 209.53 (+11\%)          & 4.94 (+28\%)         \\
LRM+Prompt-heavy & 289.01 (+53\%)          &  8.11 (+110\%)         \\ \hline
\end{tabular}
\vspace{-0.45cm}
\end{table}
%{\bf Ablation experiment:}
Fig.~\ref{fig:RD}(c) presents an ablation study of our add-on modules, LRM and the prompt generator $g_p$, for real-world noise. Note that these add-on modules together represent a two-pronged approach to joint decoding and denoising. LRM is a compressed-domain mechanism that aims to predict the latent representation of the denoised image from a noisy one; in comparison, $g_p$ is to adapt the decoding process of a pre-trained base codec. Specifically, we investigate \textcolor{black}{two LRM implementations, \textit{LRM} and \textit{LRM-light}, and} two prompt generators, \textit{Prompt} and \textcolor{black}{\textit{Prompt-heavy}}. \textcolor{black}{For prompt generators,} the former is our proposed method and the latter is its heavy variant. With \textcolor{black}{\textit{Prompt-heavy}}, the prompt generator uses convolutions instead of group convolutions, and it generates prompts for all STBs in the decoder in order to explore the full potential of adapting the decoding process only. We first note that all \textcolor{black}{seven} variants notably outperform \textit{Base} in Fig.~\ref{fig:RD}(a), showing their effectiveness in the denoising reconstruction. In addition, when enabling one module at a time, \textit{LRM} proves more effective than \textit{Prompt} or \textcolor{black}{\textit{Prompt-heavy}} for the denoising reconstruction. \textcolor{black}{\textit{LRM-light} outperforms \textit{Prompt} and shows slightly worse performance than \textit{Prompt-heavy}}. This is intuitively agreeable because the base codec is pre-trained for the standard image reconstruction. When \textit{LRM} or \textit{LRM-light} works well, the pre-trained decoder should recover a noise-free image easily. We also observe that \textit{Prompt} \textcolor{black}{and \textit{Prompt-heavy} are} able to improve further on \textcolor{black}{LRMs}. Table~\ref{tab:ablation_complexity} reports their complexity characteristics. The \textcolor{black}{combinations \textit{LRM-light} + \textit{Prompt} and } LRM + \textit{Prompt} strike a good balance between compression performance and complexity. The result justifies our design choices.

\section{Conclusion}
\label{sec:conclusion}
This work introduces a Transformer-based image compression system that is capable of switching between the standard image reconstruction and denoising reconstruction without training separate decoders. It features an LRM module in the latent space and an instance-specific prompt generator. LRM is effective in refining the latents of a noisy input image for the denoising reconstruction. Our instance-specific prompt generator further complements LRM by adapting the decoding process to suit the needs of joint decoding and denoising. Both are add-on modules that have only a modest impact on the decoder's model size and computational complexity. How to extend the proposed framework to more challenging tasks, e.g. joint decoding and de-blurring, is among our future work.

%Experimental results demonstrate that both LRM and the prompt generator are able to effectively achieve joint decoding and denoising based on a pre-trained decoder. Among them, directly refining the coded latent is more efficient than adapting the decoder. Combining LRM and the prompt generator also yields improved performance.

\bibliography{egbib}
\bibliographystyle{IEEEtran}

\end{document}